\documentclass[conference]{IEEEtran}
\IEEEoverridecommandlockouts
\usepackage{cite}
\usepackage{amsmath,amssymb,amsfonts}
\usepackage{algorithmic}
\usepackage{graphicx}
\usepackage{textcomp}
\usepackage{xcolor}
\def\BibTeX{{\rm B\kern-.05em{\sc i\kern-.025em b}\kern-.08em
    T\kern-.1667em\lower.7ex\hbox{E}\kern-.125emX}}
\begin{document}

\title{Low frequency noise in AC biased metallic tunnel junctions\\
}

\author{
\IEEEauthorblockN{Nicolas Fontaine}
\IEEEauthorblockA{%
\textit{Institut Quantique}\\
\textit{Département de Physique} \\
\textit{Université de Sherbrooke}\\
Sherbrooke, QC, Canada \\
nicolas.fontaine2@usherbrooke.ca}
\and
\IEEEauthorblockN{Alexandre Dumont}
\IEEEauthorblockA{%
\textit{Institut Quantique} \\
\textit{Département de Physique} \\
\textit{Université de Sherbrooke}\\
Sherbrooke, QC, Canada \\
alexandre.dumont3@usherbrooke.ca}
\and
\IEEEauthorblockN{Bertrand Reulet}
\IEEEauthorblockA{%
\textit{Institut Quantique}\\
\textit{Département de Physique} \\
\textit{Université de Sherbrooke}\\
Sherbrooke, QC, Canada \\
bertrand.reulet@usherbrooke.ca}
}
\maketitle

%*CRITICAL: Do Not Use Symbols, Special Characters, Footnotes, or Math in Paper Title or Abstract.
\begin{abstract}
We study the effect of an AC bias on the low frequency noise, notably $1/f^\gamma$ with $\gamma<2$,  of metal-insulator-metal tunnel junctions at room temperature. The measurement is performed in the 6Hz-100kHz frequency range with an AC excitation above 1MHz. We observe that $1/f^\gamma$ noise is dominant across our measurements though the shape of the spectra varies. The effect of the DC excitation seems to be very different on the noise generated by the junction than that of the AC excitation, thus questioning the fact that the observed noise is due to resistance fluctuations that the bias only reveals.  
\end{abstract}
%-plusieurs études en DC, pas vrm en AC
%-température ordinaire
%-effet de la fréquence du bias
%-bruit 1/f toujours dominant dans le spectre
%-forme du bruit varie en fonction des paramètres


\section{Introduction}
\label{sec:Intro}
Flicker noise is ubiquitous in electronics~\cite{Buckingham:1983,VanDerZiel:1986}.A possible origin of flicker noise in conductors comes from the fluctuations in time of the value of the resistance~\cite{Dutta:1981,Rogers:1984}. Without bias, this has negligible effect on the variance of voltage fluctuations, which is still given by Johnson-Nyquist noise $S_V=4k_BTR$ where $S_V$ is the noise spectral density of voltage fluctuations, $T$ the temperature and $R$ the resistance of the sample~\cite{Johnson:1928,Nyquist:1928}. 

In the presence of a DC current $I_{DC}$, a fluctuation of resistance $\delta R(t)$ leads to a voltage fluctuation $\delta V(t)=I_{DC}\delta R(t)$, so voltage fluctuations are given by:
\begin{equation}
    S_V(f)=S_R(f)I_{DC}^2
    \label{eq:DC}
\end{equation}
\noindent with $S_R$ the spectral density of resistance fluctuations, in $\Omega^2$/Hz. With this picture in mind, replacing the DC current by an AC one oscillating at frequency $f_0$ should lead to the appearance of noise on each side of $f_0$: the $1/f$ noise observed at low frequency should be shifted into sidebands behaving like $1/|f-f_0|$. But there should be almost no $1/f$ noise at low frequency, provided $f_0$ is large enough. More precisely, one expects the noise in the presence of an AC current of amplitude $I_{AC}$ to have a spectral density given by:
\begin{equation}
    S_V(f)=\frac12[S_R(|f-f_0|)+S_R(f+f_0))]I_{AC}^2
    \label{eq:AC}
\end{equation}
\noindent It is the goal of our experiment to study this behaviour.

\section{Experimental Setup and Measurements}
The samples we have studied are commercial Nb / Al oxide / Nb tunnel junctions. The junctions have the shape of a disc of radius $5\mu$m, i.e. have an area of $78.5\mu\mathrm{m}^2$, for a DC resistance of $R\simeq50\Omega$ and a bandwidth above 500MHz. The current-voltage characteristics of the junctions is very linear: the differential resistance varies by $\sim2$\%  within the voltage range we have explored. Below we report measurements performed on one sample of resistance $R=52\Omega$. We have measured 3 other samples that show similar behaviour.

\begin{figure}[h]
    \centering
    \includegraphics[width=0.5\textwidth]{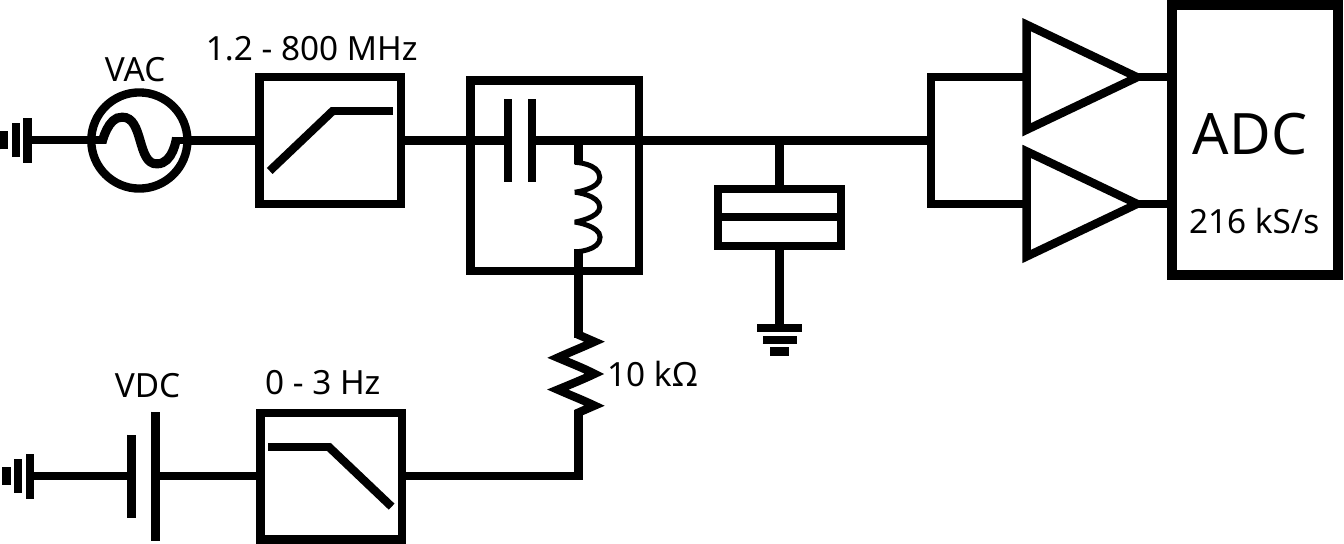}
    \caption{Experimental setup. ADC represents the two-channel digitizer.}
    \label{fig:setup}
\end{figure}

Fig. \ref{fig:setup} presents the experimental setup used to conduct our measurements. A DC- and an AC voltage sources are connected to a sample through a bias tee, after filtering to avoid the detection of possible noise generated by the sources themselves. A 10k$\Omega$ resistor on the DC line allows us to impose the current in the junction while the voltage generated by the AC source is split between the 50$\Omega$ output resistance of the source and the junction.
The voltage noise generated by the junction is amplified by two low-noise amplifiers of bandwidth 0.1Hz-1MHz, both connected to independent channels of a 24bits, 216kS/s data acquisition card. From the recorded time traces of both channels, we perform cross-correlations of the two signals in frequency domain. This allows the removal of the contribution of the voltage noise of the amplifiers (4nV$/\sqrt{\mathrm{Hz}}$ for most of the spectrum), to the total measured noise. The setup is sensitive enough to detect the thermal noise of the junctions of variance $4k_BTR$, i.e. corresponding to 0.9nV$/\sqrt{\mathrm{Hz}}$ independent of frequency, see the horizontal dashed lines in Figs. \ref{fig:bruitLowVolt} and \ref{fig:bruitHighVolt}. There is no detectable $1/f$ noise in the absence of bias.

Each spectrum is averaged over 1000 acquisitions and contains 16384 points from 6.59 to 108 000 Hz. The voltage bias is measured at the sample, after the DC and AC line attenuation. AC voltage is set to the $V_{RMS}$ equivalent of the DC values.

\section{Results}
The analysis of our results will be split into two parts: first we focus on low voltage biases ranging from 2 to 10mV, then we explore higher voltage bias, from 25 to 125mV.

\subsection{Low voltage bias}

\begin{figure}
    \centering
    \includegraphics[width=0.5\textwidth]{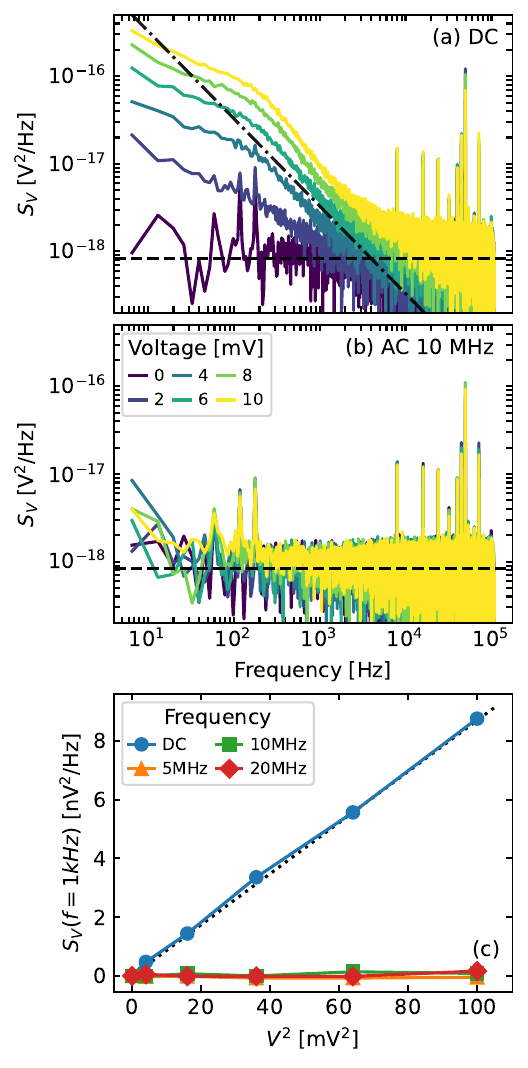}
    \caption{Results for low voltage biases. (a) Voltage noise spectral density as a function of frequency for the tunnel junction when biased by a DC excitation. The dash-dotted line shows a $1/f$ slope for reference. The dashed lines represent the expected thermal noise. (b) Voltage noise spectral density as a function of frequency for the tunnel junction when biased by an AC excitation of frequency 10 MHz. (c) Averaged noise spectral density as a function of the square of the voltage bias, DC (blue) or AC  for different frequencies. The DC and AC voltage are swept by steps of 2mV from 0 to 10mV. }
    \label{fig:bruitLowVolt}
\end{figure}

Fig. \ref{fig:bruitLowVolt} shows the results of the low bias measurements. The effect of a DC bias on the junction is reported as noise spectra in Fig. \ref{fig:bruitLowVolt} (a) for several values of the bias, from 0 to 10mV. The spectrum in the absence of bias is white, as expected for thermal noise. There is no sign of noise coming from the detection setup. The presence of a DC bias leads to an increase of noise at low frequency.  The shape of the DC spectra appears more complex than a simple power law. At frequencies below $\sim300$Hz the noise spectral density decays slower than $1/f$ (the dashed-dotted line in \ref{fig:bruitLowVolt} (a) indicates $1/f$ slope). From $\sim300$Hz to $\sim1500$Hz, $S_V(f)$ decays with a slope close to $1/f$ until it reaches the thermal noise. The noise reported here is always much greater than the shot/thermal noise of the junction $2eVR\coth{eV/2k_BT}$ which corresponds to 0.9nV$/\sqrt{\mathrm{Hz}}$.

In order to quantify the bias dependence of this noise we plot in Fig. \ref{fig:bruitLowVolt} (c) the value of the noise spectral density measured at 1kHz, $S_V(f=1\mathrm{kHz})$ averaged  over the bandwidth 0.9-1.1 kHz minus its value at equilibrium, as a function of $V^2$. We clearly observe a linear dependence, in agreement with Eq.\ref{eq:DC} derived from the idea that the noise we observe comes from the fluctuations of the resistance of the tunnel junction. We decided to plot $S_V$ as a function of the DC voltage and not the DC current as in Eq.(\ref{eq:DC}), but since the sample is almost perfectly linear, the conclusion is the same, only units change.

It is usual to characterize the amount of $1/f$ noise by the Hooge parameter $\alpha$~\cite{Hooge:1969,Hooge:1976}. In tunnel junctions the resistance of the sample is inversely proportional to the area of the junction, so the noise spectral density takes the form~\cite{Lei:2011}:
%, Eq. \ref{eq:hooge} is used to obtain the parameter at low bias values, where A represents the junction area in $\mu m^2$. In this form, the Hooge parameter has area units even if $\gamma = 1$ and $\beta = 0$, that is if the noise spectrum follows a $1/f^1$ slope and the noise level is linear with $V^2$.
%valeurs pour $S_R$, Hooge. 
\begin{equation}\label{eq:hooge}
    S_V(f)=\frac{\alpha}{Af}V^2
%    \alpha = \frac{Af^{\gamma}S_V}{V^{2+\beta}}
\end{equation}
Taking the value of $S_V$ measured at 1kHz, the dotted line of Fig.\ref{fig:bruitLowVolt}(c) corresponds to $\alpha=68\times 10^{-10}\mu\mathrm{m}^2$. Table I of \cite{Lei:2011} reports Hooge parameters of magnetic junctions using $Al_2O_3$ as insulating barrier to lie in the range 20 to 20000 $\times10^{-10}\mu \mathrm{m}^2$. In comparison, our junction can be considered quiet. However, the Hooge parameter is known to increase with the resistance-area (RA) product in magnetic tunnel junctions\cite{Nor:2006, Park:2003, Gokce:2006}. This number is a measure of the inverse transparency of the barrier. The RA product of our sample is 4.16 k$\Omega\,\mu$m$^2$, i.e. one order of magnitude below the smallest RA presented in~\cite{Lei:2011}. The comparatively high transparency of our junction may explain its relatively small Hooge parameter.
%The obtained value for the $4-10$ mV biases is  stable at $\sim 70\times 10^{-10}\mu m^2$, although the Hooge parameter can sometimes decrease as the bias voltage increases \cite{Almeida:2008}.
%, which vary from $\sim40$ to $\sim114 700$k$\Omega\,\mu$m$^2$.

We now discuss the effect of a pure AC bias on the sample. We show in Fig. \ref{fig:bruitLowVolt} (b) noise spectra for various AC voltages at an excitation frequency of $10$ MHz. In strong contrast with the results for DC bias, the AC bias has no effect on the noise spectra. Comparison between DC and AC biases are summarized in Fig. \ref{fig:bruitLowVolt} (c). AC bias has no effect for biases up to 10mV RMS and for all excitation frequencies, 5, 10 or 20 MHz. These results are in full agreement with the expectations of Eq.(\ref{eq:AC}) based on the picture of resistance fluctuations. Indeed, from Eqs.(\ref{eq:AC}) and (\ref{eq:hooge}), and the Hooge parameter mentioned above we expect a contribution of the $1/f$ noise to be below $10^{-21}\mathrm{nV}^2$/Hz for an AC voltage of 10mV at 10 MHz.

%The DC spectra show a  considerable amount of 1/f whereas the AC spectra present no noise other than the thermal noise. This is in accordance with the hypothesis presented earlier stating that the 1/f noise should be centered at the excitation frequency $f_0$. It is important to note that the maximum shot noise would amount to approximately $6.5\times10^{-23}\ V^2/Hz$, which is far below the measured noise levels for both the AC and DC biases. 

%the DC bias and Fig. \ref{fig:bruitLowVolt} (b) for the AC bias at $10$ MHz. The mean measured noise in the 900-1100 Hz bandwidth is shown for different bias values on Fig. \ref{fig:bruitLowVolt} (c). The values of Fig. \ref{fig:bruitLowVolt} (c) have had the thermal noise and shot noise subtracted and thus show only the ``excess'' 1/f noise. 

%For the DC bias, the measured 1/f noise also grows linearly with the squared voltage, as shown in Fig. \ref{fig:bruitLowVolt} (c), which again follows the expected behaviour stemming from the premise presented in Sec. \ref{sec:Intro}. Unsurprisingly, the mean noise values for the AC biases stay at zero for all frequencies.

\subsection{High voltage bias}

\begin{figure}
    \centering
    \includegraphics[width=0.5\textwidth]{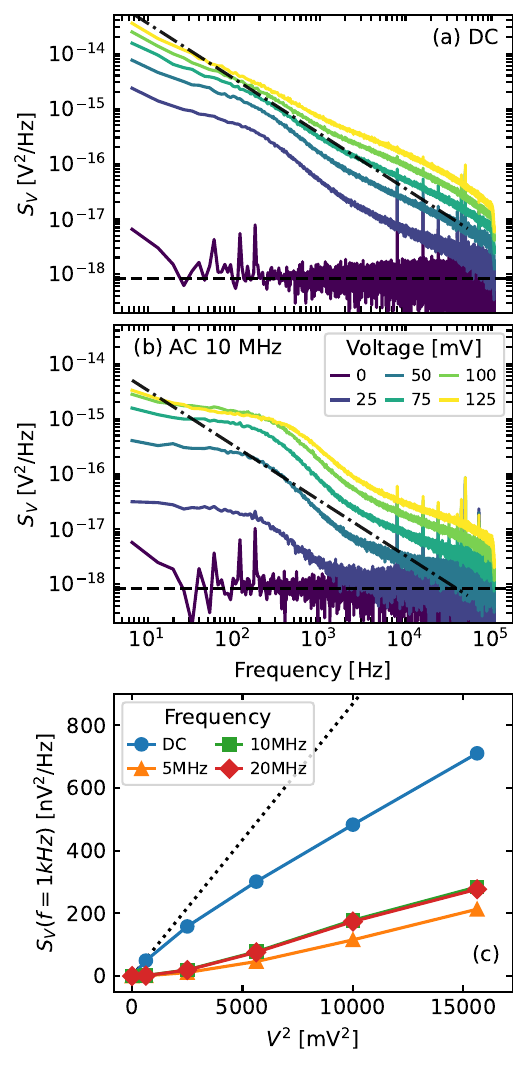}
    \caption{Results for the high voltage biases. (a) Low frequency noise spectrum of the tunnel junction when biased with a DC excitation. (b) Low frequency noise spectrum of a tunnel junction when biased with a 10 MHz excitation. The DC and AC voltage is swept by steps of 25mV from 0 to 125mV. The dash-dotted lines show a $1/f$ slope for reference. The dashed lines represent the expected thermal noise.}
    \label{fig:bruitHighVolt}
\end{figure}

We now explore the effect of DC and AC biases of higher amplitude, up to 125mV. The measurements are identical to those of the previous section, only the amplitude of the biases have increased. They are shown in Fig. \ref{fig:bruitHighVolt}.

%The results presented in this subsection concern one particular sample, but the behaviours described were observed to be comparable across multiple samples, at least as far as spectrum shape is concerned.

 Noise spectra for high DC biases are presented in Fig. \ref{fig:bruitHighVolt}(a). The shape of the DC spectra starts off similar to the low bias DC spectra, and become straighter as the bias increases. The spectrum at the highest DC bias is nearly a pure power law with a slope slightly less steep than than pure 1/f. 
 
 Noise spectra for an AC bias at 10 MHz are reported in Fig. \ref{fig:bruitHighVolt}(b). There is a clear effect of the AC excitation, in strong contrast to what happens with low AC bias.
 
 The mean measured noise in the 900-1100 Hz bandwidth is shown for different bias values in Fig. \ref{fig:bruitHighVolt}(c). Again, the values of Fig. \ref{fig:bruitHighVolt} (c) have had the thermal noise and shot noise subtracted and thus show only the excess noise. The case of DC bias is shown as blue circles. The data clearly fall below the $V^2$ law observed at low bias and indicated by a black dotted line: $S_V$ no longer scales as $V^2$ for high bias. The noise increases for AC bias: an AC excitation of 125mV RMS generates approximately a third of the noise generated by a 125mV DC voltage. There is also a small frequency dependence, the 5MHz excitation being slightly less effective than the 10 or 20MHz ones.
 
 %For the high voltage, the mean measured noise in the 900-1100 Hz bandwidth is no longer perfectly linear with the squared DC voltage and the 1/f noise under AC bias is of course now non-zero. Again, we do not presently have an explanation for these observations. The 1/f noise generated by AC bias are all lower than the noise generated with a DC bias. Interestingly, the AC excitation frequency seems to have an impact on the measured 1/f noise levels. The 5 MHz excitation leads to the lowest measured 1/f while the 10 and 20 MHz both lead to the same 1/f noise levels.

 The fact that $S_V$ does not scale as $V^2$ for DC bias and that $S_V$ is influenced by the AC bias are in strong disagreement with the idea that the $1/f$ noise is due to resistance fluctuations that do not depend on the bias. As far as we know there is no theoretical explanation for this behaviour. A possible cause could be the nonlinearity of the $I(V)$ characteristics of the junctions, although it deviates only very slightly from perfect linearity. It could also be that the resistance noise $S_R$ depends on the bias. 

%Both the DC and AC biased spectra present levels of $1/f$ noise that are far beyond the thermal and shot noise. The maximum shot noise would still amount to only approximately $8\times10^{-22}\ V^2/Hz$, which is once more well below the measured noise levels for both the DC and AC biases. The AC bias spectra are in clear disagreement with the hypothesis presented in Sec. \ref{sec:Intro} and we currently do not have an explanation for this behaviour.  Similarly to the low bias DC spectra, the AC spectra with high bias appear to have three distinct regions with clearly different slopes, although the value of these slopes differ. The first region (6-300 Hz) seems nearly flat whereas the second region (300-1500 Hz) looks to be sharper than the 1/f line. One particular thing to note is that the noise level of the first region seems to saturate between the 100 and 125mV bias. The behaviours of the spectra were homogeneous across different excitation frequency (5, 10 or 20 MHz) in our measurements.

\section{Conclusion}
We have measured the effect of DC and AC bias on the low frequency noise appearing in metallic tunnel junctions. Our result are in contradiction with the simple picture that low frequency noise is due to resistance fluctuations that do not depend on voltage. This should help shed light on the mechanism responsible for the $1/f$-like noise in these samples. Our experiment can be applied to many conducting samples that exhibit flicker noise to see if the behaviour we observe is general or not. 

\section{Acknowledgments}
We thank A. Horik and D. Grebe for their participation in the early stage of the experiment, G. Laliberté, E. Pinsolle and C. Lupien for their help with the experimental techniques, and A.-M. Tremblay for fruitful discussions. We thank Quantum e-Motion for providing us with the samples and for financial support. This work was supported by the Canada Research Chair program, the NSERC, the Canada First Research Excellence Fund, the FRQNT, and the Canada Foundation for Innovation.

% \nocite{*}
\bibliographystyle{ieeetr}
\bibliography{referencesICNF2023}

\end{document}